\documentclass[prb]{revtex4}

\usepackage{graphicx}

\begin{document}


\title{Sisyphus cooling and amplification by a superconducting qubit}


\author{M.~Grajcar$^{1,2}$, S.H.W.~van der Ploeg$^1$, A.~Izmalkov$^1$,
E.~Il'ichev$^1$, H.-G.~Meyer$^1$, A.~Fedorov$^3$, A.~Shnirman$^{4}$, and
Gerd~Sch\"on$^5$}

\affiliation{
$^1$Institute of Photonic Technology, P.O. Box 100239, D-07702 Jena,
Germany\\
$^2$Department of Experimental Physics, Comenius
University, SK-84248 Bratislava, Slovakia\\
$^3$Quantum Transport Group,
Delft University of Technology, 2628CJ Delft, The Netherlands \\
$^4$Institut f\"{u}r Theoretische Physik, Universit\"{a}t Innsbruck,
A-6020 Innsbruck, Austria\\
$^5$Institut f\"{u}r Theoretische Festk\"{o}rperphysik and DFG-Center
for Functional Nanostructures (CFN), Universit\"{a}t Karlsruhe,
D-76128 Karlsruhe, Germany
}

\begin{abstract}
Laser cooling of the atomic motion paved the way for remarkable
achievements in the fields of quantum optics and atomic physics,
including Bose-Einstein condensation and the trapping of atoms in
optical lattices. More recently superconducting qubits were
shown to act as artificial two-level atoms, displaying Rabi
oscillations, Ramsey fringes, and further quantum
effects~\cite{Nakamura99,Vion02,Chiorescu03}. Coupling
such qubits to
resonators~\cite{WallraffQED,Chiorescu04,Ilichev03,Blais04} brought
the superconducting circuits into the realm of quantum electrodynamics
(circuit QED). It opened the perspective to use superconducting  
qubits as
micro-coolers or to create a population inversion in the qubit to  
induce lasing behavior of the resonator~\cite 
{Martin04,Rabl04,Niskanen07,Hauss07}. Furthering
these analogies between quantum optical and superconducting systems we
demonstrate here Sisyphus cooling~\cite{Wineland92} of a low frequency
LC oscillator coupled to a near-resonantly driven superconducting qubit.
In the quantum optics setup the mechanical degrees of freedom of  
an atom are cooled by laser driving the atom's electronic degrees of  
freedom.
Here the roles of the two degrees of freedom are played by the LC  
circuit and the qubit's levels, respectively.
We also demonstrate the
counterpart of the Sisyphus cooling, namely Sisyphus amplification.

For red-detuned high-frequency driving of the qubit the low-frequency  
LC circuit performs work in the forward and backward part of the  
oscillation cycle, always pushing the qubit up in energy, similar to  
Sisyphus
who always had to roll a stone uphill. The oscillation cycle is  
completed with a relaxation process, when the work performed by the  
oscillator together with a quantum of energy of the high-frequency  
driving
is released by the qubit to the environment via spontaneous emission.
For blue-detuning the same mechanism creates
excitations in the LC circuit with a tendency towards lasing and the
characteristic line-width narrowing. In this regime ``lucky
Sisyphus'' always rolls the stone downhill. Parallel to the
experimental demonstration we analyze the system theoretically and
find quantitative agreement, which supports the interpretation and  
allows us to estimate  system parameters.
\end{abstract}

\maketitle

\begin{figure}
\includegraphics[width=8cm]{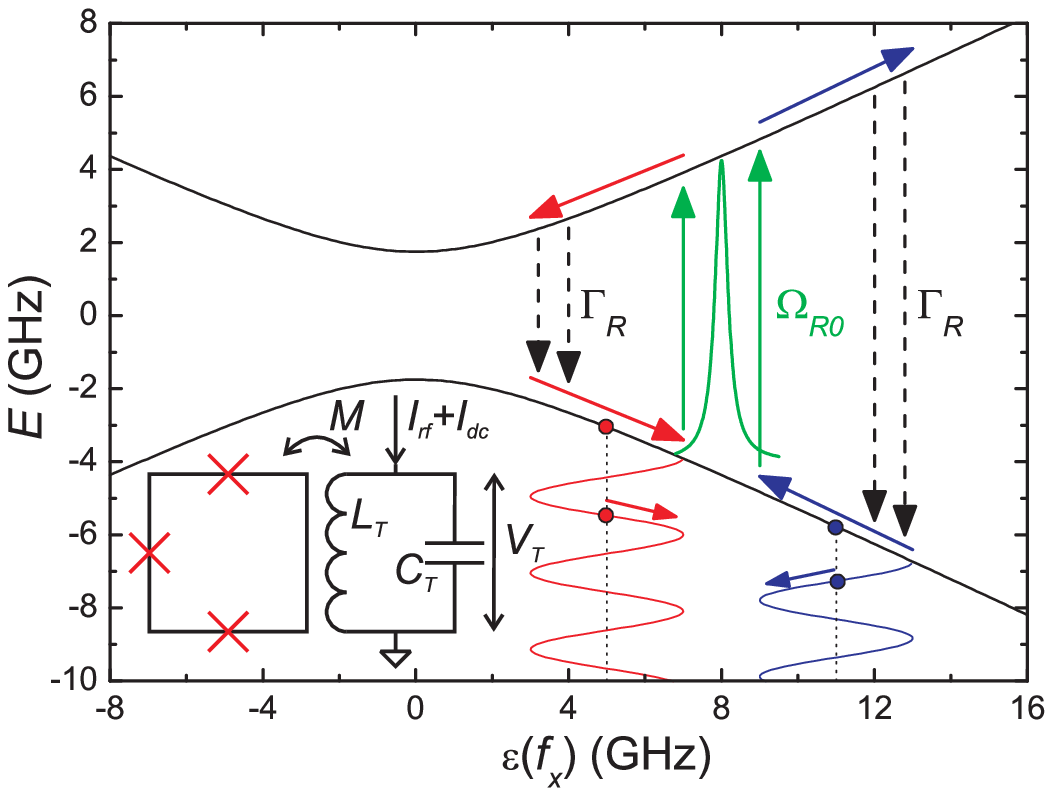}
\includegraphics[width=3.5cm]{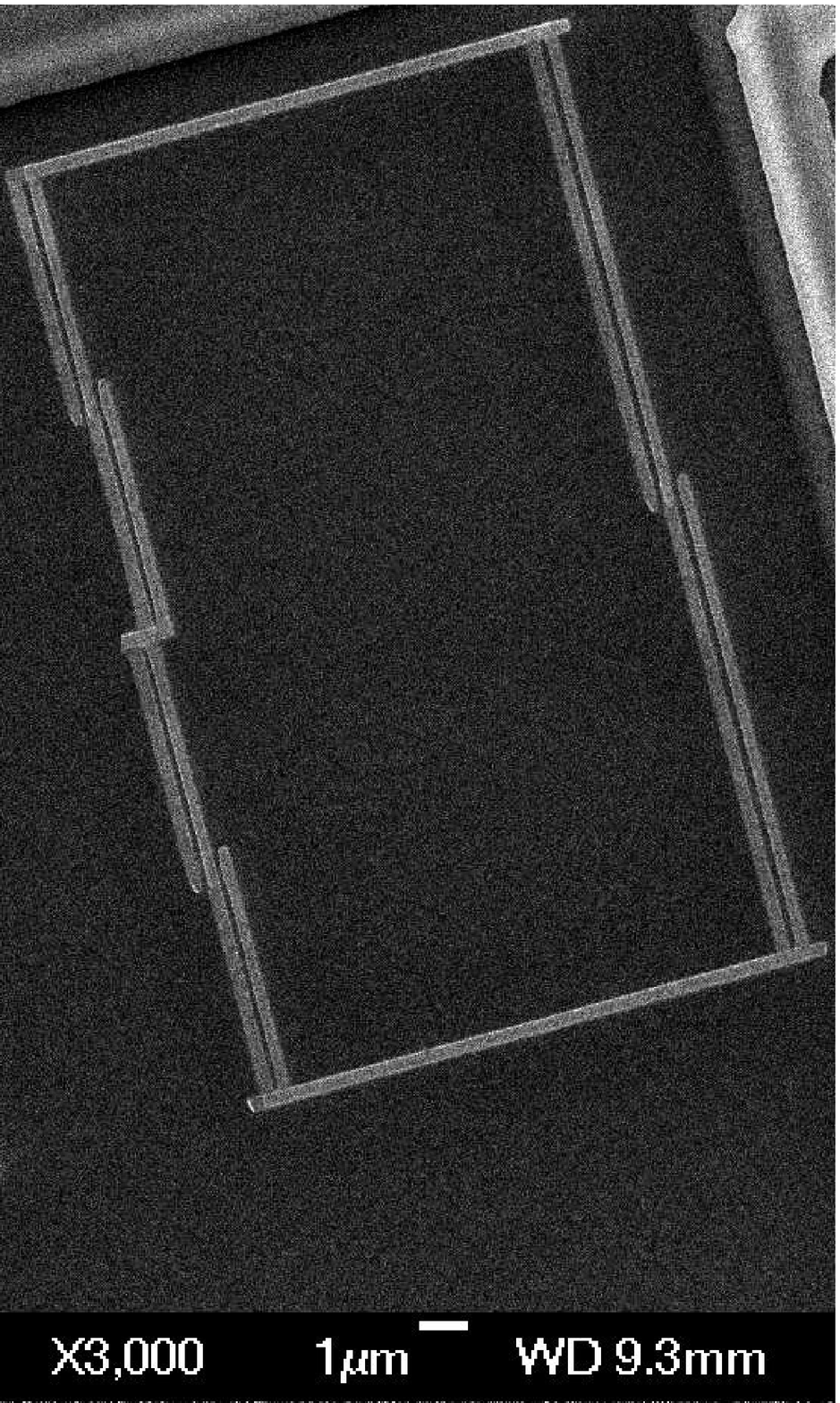}
\caption{\textbf{(a)} The energy levels of the qubit as a function of  
the energy
bias of the qubit $\varepsilon(f_x)=2\Phi_0 I_p f_x$. The sinusoidal
current in the tank coil, indicated by the wavy line, drives the
  bias of the qubit. The starting point of the cooling
(heating) cycles is denoted by blue (red) dots. The resonant
excitation of the qubit due to the high-frequency driving,  
characterized by $\Omega_{R0}$, is
indicated by two green arrows and by the Lorentzian depicting the
width of this resonance. The relaxation of the qubit is denoted by
the black dashed arrows. The inset shows a schematic of the qubit
coupled to an LC circuit. The high frequency driving is provided by
an on-chip microwave antenna.
\textbf{(b)} SEM picture of the superconducting
flux qubit prepared by shadow evaporation technique.} \label{Fig:En_lev}
\end{figure}

The system considered is shown in the inset of
Fig.~1. It consists of a three-junction flux qubit~\cite{Mooij99},  
with the two qubit states
corresponding to persistent currents of amplitude $I_p$ flowing
clockwise and counterclockwise. When operated in the vicinity of
the degeneracy point, $f_x \equiv \Phi_\mathrm{x}/\Phi_0-1/2
\approx 0$, where $\Phi_\mathrm{x}$ is the magnetic flux applied to
the qubit loop and $\Phi_0=h/2e$ the flux quantum, the Hamiltonian
of the qubit in the basis of the persistent current states reads
\begin{equation}
H=-\frac{1}{2}\varepsilon(f_x) \sigma_{z} -\frac{1}{2} \Delta
\sigma_{x}\ . \label{eq:Hamiltonian_classical}
\end{equation}
Here $\sigma_x,\sigma_z$ are Pauli matrices, $\Delta$ is the
tunneling amplitude, and $\varepsilon(f_x)=2\Phi_0 I_{p} f_{x}$ is
the energy bias. The energy levels
of the isolated qubit, separated by
$
\Delta E(f_{x})=\sqrt{\varepsilon^2(f_{x})+\Delta^2}
$
are shown in Fig.~1.
The qubit is driven by a high-frequency field with frequency $\omega_d 
$ and
coupled via a mutual inductance $M$ to a low-frequency tank circuit  
with frequency
$\omega_T \simeq 2\pi \times 20\ \mathrm{MHz} \ll \Delta/\hbar$.
Both
circuit to the qubit
can be included at this stage via their contributions to the external  
flux
$\Phi_x$. Since the eigenfrequency of the tank circuit, $\omega_T$,
is much lower than the level spacing of the qubit, the oscillations
of the current in the tank circuit can be treated in an adiabatic
approximation, i.e., the current in the tank circuit shifts the bias
flux of the qubit by $\Phi_x(t) = M[I_{dc}+I_{\rm rf}(t)]$.

The system mimics the Sisyphus mechanism of damping (cooling) and
amplification (heating) known from quantum
optics~\cite{Wineland92}. This mechanism is illustrated in
Fig.~1. Here we describe the damping (cooling, hence marked in blue) for a
situation where the driving is red-detuned, $\hbar \omega_d <
\Delta E$; the amplification (marked in red) for blue-detuning can be described in
an analogous way. The oscillations of the current in the tank
circuit, $I_{\rm rf}(t)$, lead to oscillations of $\varepsilon(f_x)$  
around
a value determined by the $dc$ component, $I_{dc}$. In the first
part of the cycle, when the qubit is in the ground state, the
current shifts the qubit towards the resonance, $\Delta E
=\hbar\omega_d$, i.e., the energy of the qubit grows due to work
  done by the LC circuit. Once the system reaches the vicinity of
the resonance point, the qubit can get excited, the energy being
provided by the high-frequency driving field. With parameters
adjusted such that this happens at the turning point of the
oscillating trajectory, the qubit in the excited state is now shifted
by the current away from the resonance, such that the
qubitÇs energy continues to grow. Again the work has to be provided  
by the LC
circuit. The cycle is completed by a relaxation process which takes  
the qubit back to the ground state. The maximum effect is achieved when
the driving frequency and relaxation rate are of the same order of
magnitude. If the relaxation is too slow the state is merely shifted
back and forth adiabatically during many periods of oscillations.
Note that the complete cycle resembles the ideal Otto-engine  
thermodynamic cycle~\cite{Quan07}.

Two types of measurements have been performed on the system. In the  
first the
LC tank circuit is additionally driven near-resonantly by a low- 
frequency $rf$ current,
and the response of the LC circuit is detected using lock-in  
techniques. In these
measurements we identify the influence of the high-frequency driven  
qubit on the effective quality factor and  eigenfrequency of the tank  
circuit. We associate a reduction of the effective quality factor with  
cooling and identify regions in parameter space where the effect is  
optimized.

In the second type of measurements the low-frequency $rf$-driving is  
switched off, while the emission of the LC tank circuit is monitored  
by a spectrum analyzer.
This analysis probes the influence of the qubit on the
effective quality factor, and, most importantly, allows us to  
determine the energy stored in the tank circuit, i.e., its effective  
temperature. Thus we are able to demonstrate cooling or heating of  
the LC oscillator.

We start with a qualitative analysis of the first type of measurement.
The LC tank circuit is driven by a current source with amplitude $I_ 
{T}^{d}$.
Then
the amplitude, $V_{T}$, of the voltage
oscillations across the tank circuit, which is measured in the
experiment, is
\begin{equation}
     V_T= \omega_T L_T I_{T}= Q \, \omega_T L_T I_{T}^{d}\ ,
     \label{Eq:V}
\end{equation}
where  $I_{T} = \sqrt{2 \langle I_{\rm rf}^2\rangle} = Q I_{T}^{d}$  
is the actual amplitude of the $rf$ current in the inductance,
and $Q$ is the effective quality factor of the tank circuit.
It is given by the ratio, $Q= 2\pi W_T/A$, between the energy stored  
in the tank,
$W_T=L_T I_{T}^2/2$, and energy loss per period $A$.
The latter is composed of two contributions, $A
= A_T + A_{\rm Sis}$. The intrinsic losses of the tank are given by  
$A_T=2\pi W_T/Q_0$, where $Q_0$ is the
intrinsic quality factor of the tank circuit. The average work
done by the tank on the qubit in one period, $A_{\rm
Sis}$, can be estimated as follows:
We consider the optimal situation when
the oscillator brings the qubit into the
resonance at the turning point of its trajectory (Fig.~1).
We assume further that the state of the qubit is
instantaneously ``thermalized'', with equal probabilities to remain
in the ground state or to get excited. In the latter case, after  
leaving the
area of resonance the qubit can relax with rate $\Gamma_R$.
The probability of qubit relaxing during the time $0 < t < T$ (here  
$T=2\pi/\omega_{T}$)
within an interval $dt$ is given by $dP = \exp(-\Gamma_R t)\,\Gamma_R  
dt$.
Thus we obtain the
average value of the work (taking into account the probability $1/2$
of the initial state to be excited)
\begin{eqnarray}
  A_{\rm Sis}&=&MI_pI_{T}\int dP \,[1-\cos(\omega_{T} t)]\nonumber\\
&=& MI_pI_{T} f(\Gamma_R,\omega_{T})\ ,
  \label{Eq:A_q}
\end{eqnarray}
where
\begin{equation}
f(\Gamma_R,\omega_{T}) \equiv \left(1-e^{-2\pi\Gamma_R
/\omega_{T}}\right)\,\frac{\omega_{T}^2}
{\omega_{T}^2+\Gamma_R^2}\ .
\end{equation}
This function demonstrates that the optimal situation for damping or  
amplification is
reached when $\Gamma_R \sim \omega_{T}$. For the effective quality
factor we obtain
\begin{eqnarray}
Q=Q_0 \left(1\pm\frac{A_{\rm Sis} Q_0}{2\pi W_T}\right)^{-1}=
Q_0\left(1\pm \frac{M I_p Q_0 \omega_{T}
f(\Gamma_R,\omega_{T})} {\pi V_T}\right)^{-1} \label{Eq:QQ0}\ ,
\end{eqnarray}
where $\pm$ stand for the damping/amplification. Substituting into
Eq.~(\ref{Eq:V}) we arrive at
\begin{equation}
V_T-V_{T0}= \mp M \omega_{T} I_p Q_0 f(\Gamma_R,\omega_{T})/\pi
\ , \label{Eq:dV_T}
\end{equation}
where $V_{T0}=Q_0\omega_T L_T I_{T}^{d}$ is the voltage on the tank
circuit far from the resonance. Eq.~(\ref{Eq:dV_T}) provides the
estimate for the increase/decrease of the tank voltage for  large
driving currents $I_{T}^{d}$, such that the qubit spends most of the
time away from the resonant excitation area. For weaker driving,
when the times spend away and within the excitation area are
comparable, the damping effect becomes weaker.

Our results of the first type of experiment are shown in Fig.~2.
The dips (peaks) correspond to the Sisyphus damping (amplification)
of the tank circuit, i.e., to the decrease (increase) of the  
effective quality factor $Q$.
The central dip is due to the shift of the
oscillator frequency when the qubit is at it's degeneracy
point \cite{Greenberg02a} and is not related to the Sisyphus effect.
We observe from Fig.~2b that the additional
voltage saturates for large driving amplitudes at approximately $200 
$~nV.
Using this value and Eq.~(\ref{Eq:dV_T}) we see that indeed the
relaxation time is close to the period of oscillations, $1/\Gamma_R
\approx 0.95 T$.
\begin{figure}
\includegraphics[width=7.5cm]{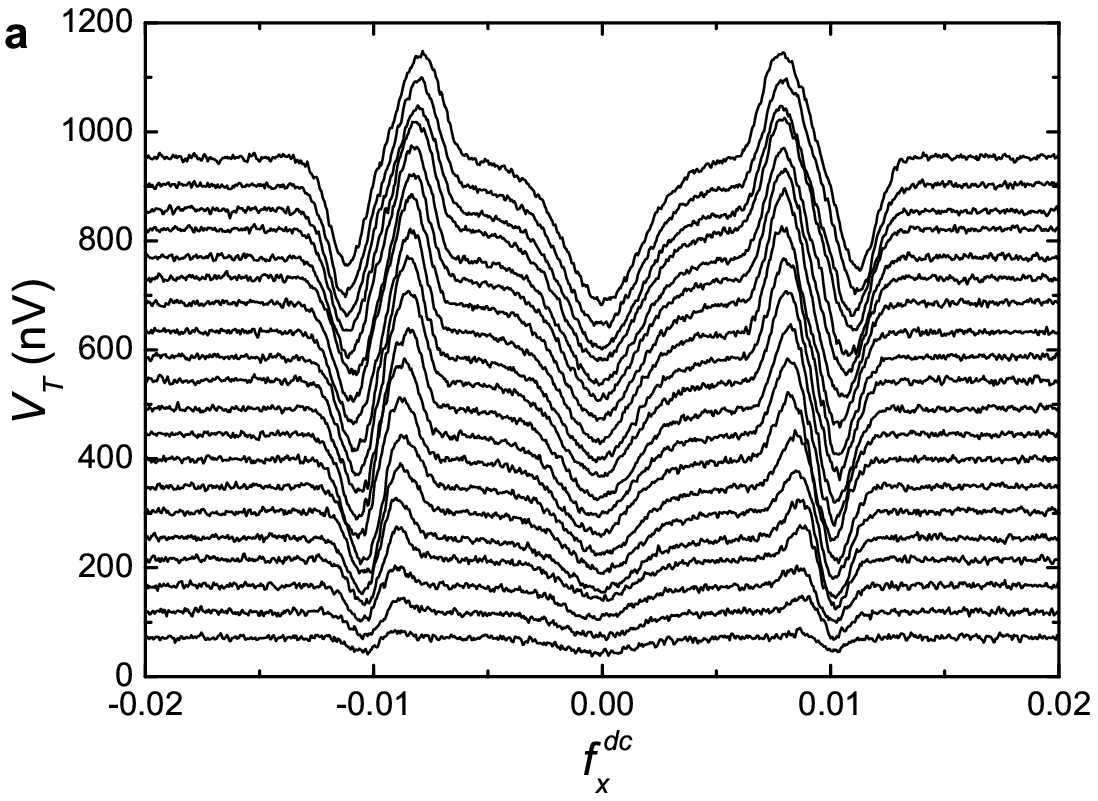}
\includegraphics[width=7.5cm]{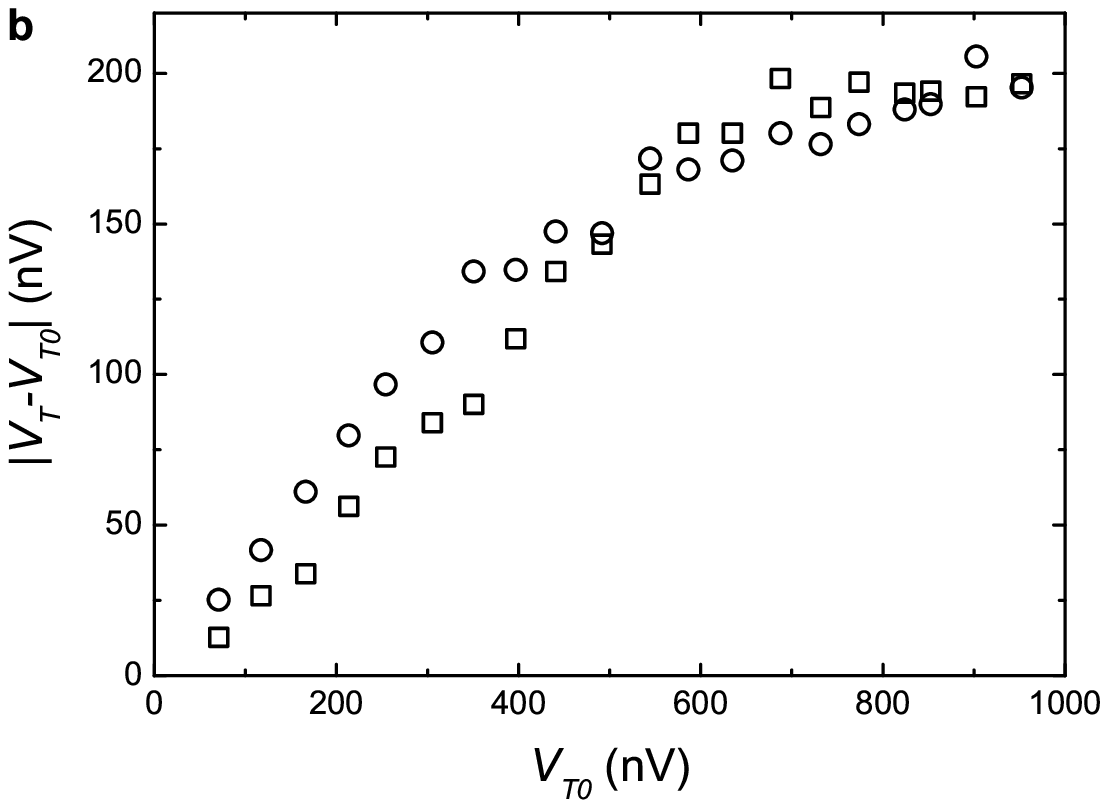}
\caption{\textbf{(a)} Amplitude of the $rf$ tank voltage as a
function of the $dc$ magnetic bias of the qubit, $f_{x}^{dc}$,
for a microwave driving frequency $\omega_{d}=2\pi\times 14.125$ GHz
and
various amplitudes of the $rf$ current driving the tank circuit.
  \textbf{(b)} The
height of the peaks (solid squares) and dips (solid circles) as a
function of the  voltage amplitude of the unloaded tank circuit $V_ 
{T0}$. The
voltage $V_{T0}$ is equal to, e.g., $V_T$ taken at $f_{x}^{dc}=0.02$
where the effect of the interaction between the tank circuit and the
qubit is negligible. The height saturates
near $200$~nV. \label{Fig:spectr_ampl} }
\end{figure}

The results for the second type of experiment are shown in Fig.~3.
The comparison shows that at the bias
point corresponding to maximum Sisyphus damping the resonant line
widens in good agreement with the results for the driven LC circuit.
At the bias point corresponding to maximum amplification the
line narrows, showing a tendency towards lasing behavior. We can
extract the quality factors in these two regimes, as well as for the
bias point far from the resonance where no additional damping
occurs. The comparison of the experimental results with
Eq.~(\ref{Eq:QQ0}) yields a quantitatively reasonable agreement.

\begin{figure}
\includegraphics[width=7.5cm]{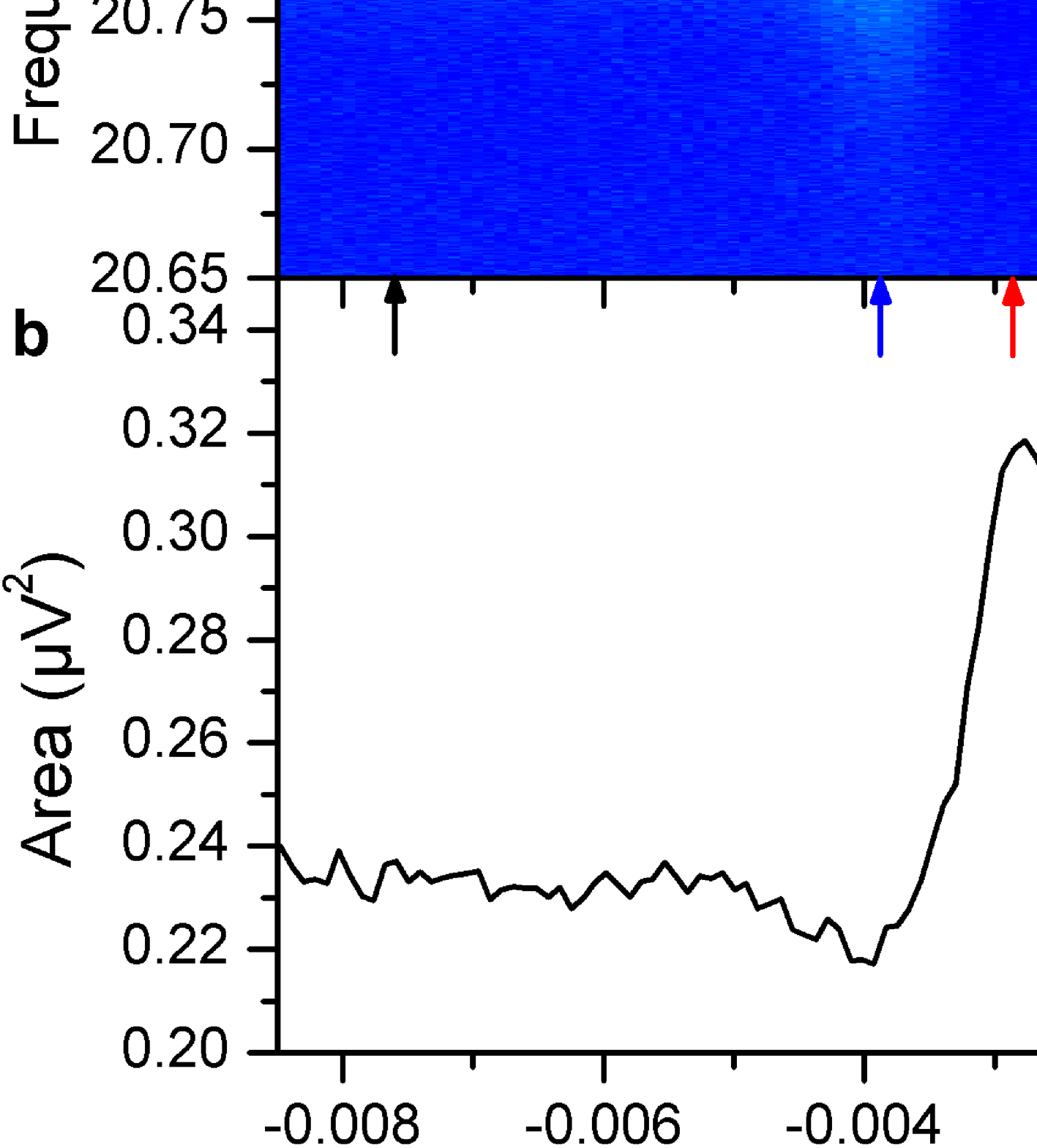}
\includegraphics[width=7.5cm]{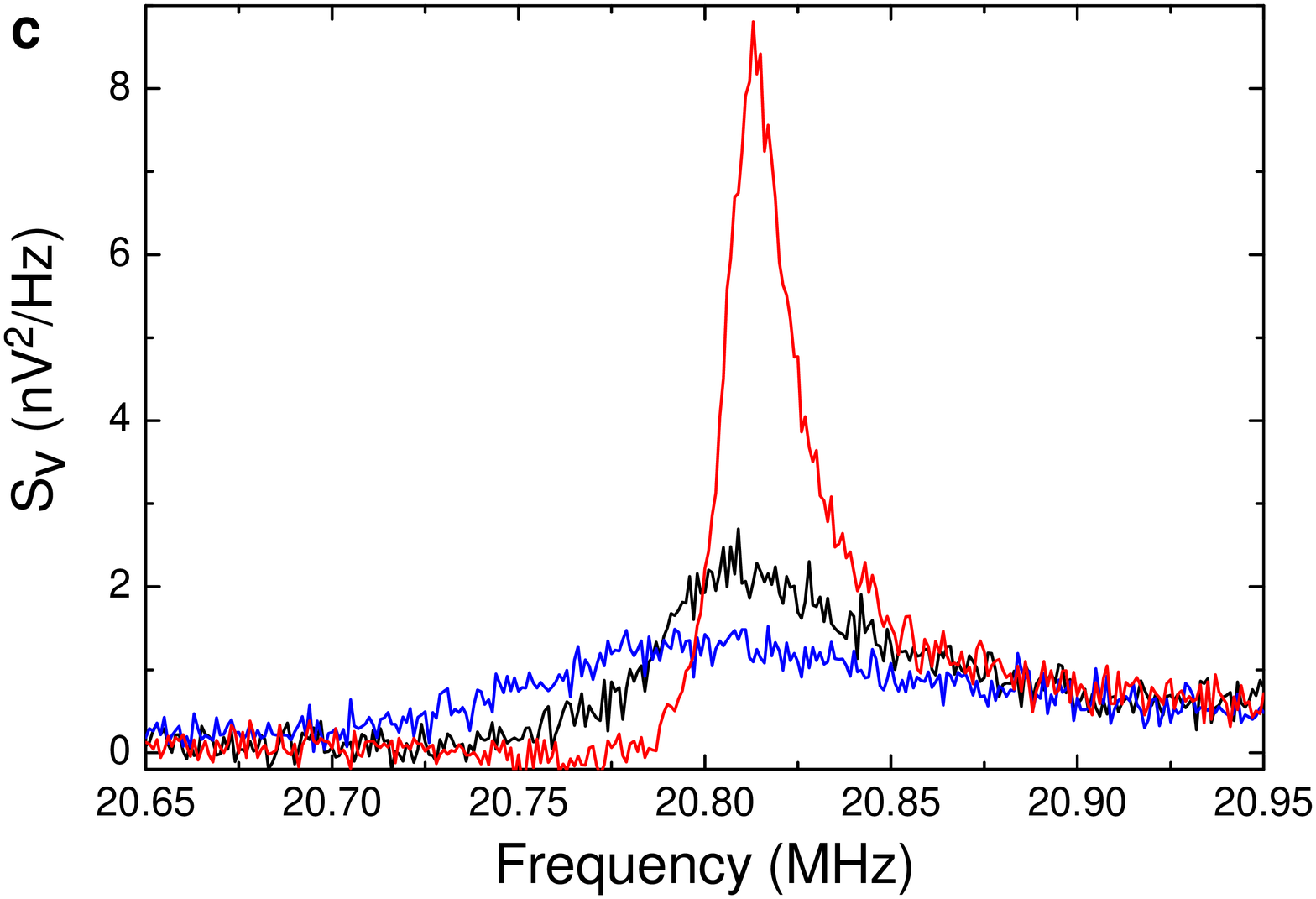}
\caption{ \textbf{(a)} Spectral density of the voltage noise in the  
LC circuit,
$S_V(\omega)$,
measured when the low-frequency $rf$-driving is switched off while  
the high-frequency qubit driving
is fixed at  $\omega_{d}=2\pi\times 6$~GHz.
$S_V$ is shown as a function of frequency (vertical axis) and
normalized magnetic flux in the qubit $f_x^{\rm dc}$ (horizontal axis).
\textbf{(b)} The integral, $\int S_V(\omega)d\omega$, evaluated using  
the data from panel \textbf{(a)} as a function of $f_x^{\rm dc}$. The  
integral is directly proportional to
the number of quanta (effective temperature) of the tank circuit. At  
optimal dc
bias $f_x^{\rm dc}$ the effective temperature $T$ of the oscillator  
is lower
than the temperature $T_0$ away from the resonance, $(T_0-T)/T_0=0.08 
$, corresponding to an $8\%$ cooling.
  \textbf{(c)} Spectral density $S_V(\omega)$ measured at three  
different values of $f_x^{\rm dc}$ corresponding to
damping (blue), amplification (red), and away from the resonance  
(black).
These values of $f_x^{\rm dc}$ are marked, respectively, by
blue, red, and black arrows in  panel \textbf{(a)}.
}
\label{Fig:Sv}
\end{figure}

Finally, the results displayed in Fig.~3 allow us to estimate
the efficiency of the cooling or heating in the two regimes.
Integrating the power spectra we observe that in the damping regime
the number of photons in the LC circuit decreases by about $8\%$
compared to the undamped case. The cooling effect was observed also  
for other
microwave frequencies and lies in the range of $6\%-13\%$.
Thus there is only little cooling. One reason is that our system is  
optimized
towards maximum Sisyphus damping rather than minimum cooling  
temperature.
Indeed, damping is optimized when the resonant point is reached at  
the turning point
of the oscillators trajectory. In this regime even a small reduction  
of the oscillators
amplitude, i.e., a weak cooling, suffices to switch the whole  
mechanism off.
We expect to be able to optimize in future
experiments  the system towards efficient cooling as
well.

To develop the discussion further we model the Sisyphus damping as an  
additional effective bath coupled to the oscillator, which we  
characterize by
a quality factor $Q_{\rm Sis}$ and temperature $T_{\rm Sis}$.
The standard analysis gives the total quality factor and temperature,
$Q^{-1} = Q_0^{-1} + Q_{\rm Sis}^{-1}$ and
$T Q^{-1} = T_0 Q_0^{-1} + T_{\rm Sis} Q_{\rm Sis}^{-1}$, respectively.
Here $T_0$ is the
temperature of the oscillator without Sisyphus damping, which in our  
case
is determined mostly by the amplifier. Our data show that $Q \sim  
Q_0/2$. Thus
the coupling strength of the Sisyphus mechanism is comparable to that
of the rest of the environment, i.e., $Q_{\rm Sis} \approx Q_0  
\approx 340$.
In the present experiment it looks as if $T_{\rm Sis}$
were comparable with $T_0$ and we have only little cooling.

In contrast, in the amplification mode of our experiment
the line gets much sharper, which formally can be expressed
by choosing $Q_{\rm Sis}$ negative, $Q_{\rm Sis} \approx - 600$.
This means that if no other bath was present the system would show
the lasing instability.
Consistently, the Sisyphus mechanism, acting as an active medium with
population inversion, is described by a negative temperature $T_{\rm  
Sis}$.
Its value is not known, but we can get a lower bound for the effective
temperature from the inequality $T/Q > T_{0}/Q_{0}$. It
predicts an increase in temperature and number of photons by a
factor $\sim 2$. The experiment shows  a $36\%$ increase, which
demonstrates the expected trend, even though some quantitative
discrepancy remains.

We now outline the theoretical analysis of the problem.
Quantizing the oscillations in the tank circuit we arrive at the  
Hamiltonian
\begin{eqnarray}
\label{eq:Hamiltonian_quantum} H & = &
-\frac{1}{2}\,\varepsilon\left(f_{x}^{dc}\right){\sigma}_{z}
-\frac{1}{2}\, \Delta {\sigma}_{x} -
\hbar\Omega_{R0}\cos(\omega_{d}t)\, {\sigma}_{z}\nonumber\\
  &  &+\ \hbar\omega_{T}\,{a}^{\dagger}
{a} \ +\ g\, {\sigma}_{z}\left({a}+{a}^{\dagger}\right)\ .
\label{H1}
\end{eqnarray}
where the coupling constant is $g = M I_p I_{T,0}$, and $I_{T,0}=\sqrt 
{\hbar\omega_{T}/2L_{T}}$
is the amplitude of the vacuum fluctuation of the current in the
LC oscillator.
The third term of Eq.~(\ref{eq:Hamiltonian_quantum}) describes
the high-frequency driving with amplitude $\Omega_{R0}$.
After transformations
to the eigenbasis of the qubit and some approximations appropriate  
for the considered situations (and described in the Appendix) the  
Hamiltonian reduces to
\begin{eqnarray}
\label{eq:H_RWA} {H} & = & -\frac{1}{2}\,\Delta E\,{\sigma}_{z}
+\hbar\Omega_{R0}\cos\left(\omega_{d}t\right)
\cos\zeta\,{\sigma}_{x}
+\hbar\omega_{T}\, {a}^{\dagger}{a}\nonumber\\
&+&g\sin\zeta\,{\sigma}_{z}\left({a}+{a}^{\dagger}\right)
-\frac{g^{2}}{\Delta E}\cos^{2}\zeta\,{\sigma}_{z}
\left({a}+{a}^{\dagger}\right)^{2}\ ,
\label{H2}
  \end{eqnarray}
with $\Delta E \equiv \sqrt{\epsilon^2+\Delta^2}$ and $\tan\zeta\equiv 
\varepsilon/\Delta$.
Thus we obtain the effective Rabi frequency $\Omega_{R0}\cos\zeta$  
and the effective constant of
qubit-oscillator linear coupling $g\sin\zeta$. As we need both these  
terms for the Sisyphus cooling
the qubit should be biased neither at the symmetry point nor very far  
from it.
The second order term $\propto g^2$ is responsible, e.g., for the  
qubit-dependent shift of the oscillator
frequency~\cite{Greenberg02a}.

To account for the effects of dissipation we consider the Liouville  
equation for the density operator of the system including the two  
relevant damping terms,
\begin{equation}\label{Eq:master} \dot \rho=-\frac{i}{\hbar}\left[H,
\rho\right]+L_Q\, \rho+L_R\,\rho\ .\end{equation} As far as the
qubit is concerned we consider spontaneous emission with rate
$\Gamma_R$ and pure dephasing with rate $\Gamma_\varphi^*$. Hence, we  
have
\begin{eqnarray}
L_Q\, \rho&=&\frac{\Gamma_R}{2}\left(2 \sigma_- \rho \sigma_+ -
\rho\sigma_+\sigma_- -\sigma_+\sigma_-\rho \right)\nonumber\\
&+&\frac{\Gamma_{\varphi}^*}{2} \left(\sigma_z\rho\sigma_z -
\rho\right)\ .
\end{eqnarray}
We neglect the excitation rate since the qubit's energy splitting
exceeds the temperature. Thus the standard longitudinal relaxation
time $T_1$ is given by $T_1^{-1} = \Gamma_R$. The rates
$\Gamma_R$ and $\Gamma_\varphi^{*}$ may depend on the working point,
i.e., on $\zeta$, which in turn is determined by the driving  
frequency $\omega_{d}$ by the condition $\hbar \omega_{d}\approx  
\Delta E$. The rates
should describe the dissipation around these values of $\zeta$.
It should be mentioned that pure
dephasing is frequently caused by the $1/f$ noise~\cite{Ithier05}, for which the
Markovian description is not applicable. We hope, however, that the
main features are still captured provided $\Gamma_\varphi^*$ is
chosen  properly.

The resonator damping term can be written in usual form~\cite{Gardiner},
\begin{eqnarray}
&&L_R \, \rho=\frac{\kappa}{2}\left(N_{\rm
th}+1\right)\left( 2a \rho a^{\dagger} -a^{\dagger} a\rho -\rho
a^{\dagger}a \right)\nonumber\\
&&+\frac{\kappa}{2}N_{\rm th}\left( 2
a^{\dagger} \rho a-a a^{\dagger}\rho -\rho aa^{\dagger} \right),
\end{eqnarray}
where $\kappa=\omega_T/Q_0$ characterizes the strength of the  
resonator damping,
and $N_{\rm th}=\left[\exp(\hbar\omega_T/k_B T)-1\right]^{-1}$  is  
the thermal
average number of photons in the resonator.
We solve the master equation (\ref{Eq:master}) numerically in the
quasi-classical limit, i.e., for $n\equiv \langle a^{\dag}a
\rangle \gg 1$, and for various choices of the unknown parameters
$\Gamma_R$, $\Gamma_\varphi^{*}$, and $\Omega_{R0}$.
As the system is harmonically driven, we determine the response
of the observables/density matrix at the driving frequency, and,
finally, find the amplitude of the driven voltage oscillations
across the tank circuit.
\begin{figure}
\includegraphics[width=8.5cm]{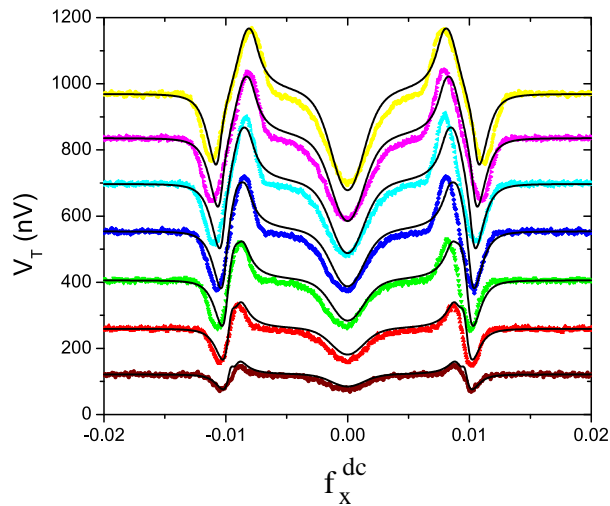}
\caption{Color lines: experimental data. Black lines: numerical
solution of Eqs.~(\ref{Eq:master}) for $\Omega_{R0}=2\pi\times
0.5$~GHz, $\Gamma_R=1.0\cdot 10^8 s^{-1}$,
$\Gamma_\varphi^{*}=5.0\cdot 10^9 s^{-1}$. The central dip is due to
the quadratic coupling term in (\ref{eq:H_RWA}) which causes a shift
of the oscillator frequency~\cite{Greenberg02a}.} \label{Fig:fit2}
\end{figure}
As shown in Fig.~4 by fitting the system parameters within a  
reasonable range
we reproduce well the experimental findings.
We should mention that a good fit can be obtained
for a relatively wide range of values of the Rabi frequency
$\Omega_{R0}$ and pure dephasing rate $\Gamma_\varphi^{*}$, provided
we keep their product roughly constant. On the other hand,
$\Gamma_R$ is determined rather accurately.

The Sisyphus damping by a superconducting qubit analyzed in this  
paper can also be
applied to nanomechanical systems~\cite 
{naik-2006-443,blanter-2004-93,blencowe-2005-7,bennett-2006-74}.
With proper optimization
towards minimum cooling temperature one could cool a mechanical
resonator which would enable, e.g., ultra-sensitive detection  
applications
like MRFM~\cite{MRFM} with quantum limited precision.

\ \\


M.G. was supported by Grants VEGA 1/0096/08 and APVV-0432-07.
We further acknowledge the financial support from the EU (RSFQubit  
and EuroSQIP).

Correspondence and requests for materials
should be addressed to M.G.~(email: grajcar@fmph.uniba.sk).

\newpage
{\bf Appendix}

After transformation to the eigenbasis of the qubit the
Hamiltonian (\ref{H1}) becomes
\begin{eqnarray}
\label{eq:H_eigenbasis} {H} & = & -\frac{1}{2}\, \Delta
E\,{\sigma}_{z}- \,\hbar\Omega_{R0}\cos\left(\omega_{d}t\right)
  \left(\sin\zeta\,{\sigma}_{z}
-\cos\zeta\,{\sigma}_{x}\right)\nonumber\\
  & + & \hbar\omega_{T}\,{a}^{\dagger}{a}\
+\, g\left(\sin\zeta\,{\sigma}_{z}-
  \cos\zeta\,{\sigma}_{x}\right)\left({a}
+{a}^{\dagger}\right) \nonumber\ ,
\end{eqnarray}
with $\tan\zeta=\varepsilon/\Delta$ and $\Delta E \equiv
\sqrt{\varepsilon^2 + \Delta^2}$.
Because of the large difference
of the energy scales between the qubit and the oscillator, $\Delta
E \gg \hbar \omega_{T}$, we can drop within the usual
rotating wave approximation (RWA) the
longitudinal driving term
$-\hbar\Omega_{R0}\cos\left(\omega_{d}t\right)\sin\zeta\,{\sigma}_{z}$.
On the other hand, we retain the transverse coupling term,
$-g\cos\zeta\,{\sigma}_{x}\left({a} +{a}^{\dagger}\right)$,
but transform it by
employing a Schrieffer-Wolff transformation,
${U}_{S}=\exp\left(i{S}\right)$, with generator ${S}=(g/\Delta
E)\cos \zeta\,\left({a}+{a}^{\dagger}\right)\,{\sigma}_{y}$, into
a second order longitudinal coupling.
The Hamiltonian then reduces to the form (\ref{H2}).


\begin{thebibliography}{10}
\expandafter\ifx\csname url\endcsname\relax
  \def\url#1{\texttt{#1}}\fi
\expandafter\ifx\csname urlprefix\endcsname\relax\def\urlprefix{URL }\fi
\providecommand{\bibinfo}[2]{#2}
\providecommand{\eprint}[2][]{\url{#2}}

\bibitem{Nakamura99}
\bibinfo{author}{Nakamura, Y.}, \bibinfo{author}{Pashkin, Y.~A.} \&
  \bibinfo{author}{Tsai, J.~S.}
\newblock \bibinfo{title}{Coherent control of macroscopic quantum states in a
  single-cooper-pair box}.
\newblock \emph{\bibinfo{journal}{Nature}} \textbf{\bibinfo{volume}{398}},
  \bibinfo{pages}{786--788} (\bibinfo{year}{1999}).

\bibitem{Vion02}
\bibinfo{author}{Vion, D.} \emph{et~al.}
\newblock \bibinfo{title}{Manipulating the quantum state of an electrical
  circuit}.
\newblock \emph{\bibinfo{journal}{Science}} \textbf{\bibinfo{volume}{296}},
  \bibinfo{pages}{886--889} (\bibinfo{year}{2002}).

\bibitem{Chiorescu03}
\bibinfo{author}{Chiorescu, I.}, \bibinfo{author}{Nakamura, Y.},
  \bibinfo{author}{Harmans, C.} \& \bibinfo{author}{Mooij, J.}
\newblock \bibinfo{title}{Coherent quantum dynamics of a superconducting flux
  qubit}.
\newblock \emph{\bibinfo{journal}{Science}} \textbf{\bibinfo{volume}{299}},
  \bibinfo{pages}{1869--1871} (\bibinfo{year}{2003}).

\bibitem{WallraffQED}
\bibinfo{author}{Wallraff, A.} \emph{et~al.}
\newblock \bibinfo{title}{Strong coupling of a single photon to a
  superconducting qubit using circuit quantum electrodynamics}.
\newblock \emph{\bibinfo{journal}{Nature}} \textbf{\bibinfo{volume}{431}},
  \bibinfo{pages}{162--167} (\bibinfo{year}{2004}).

\bibitem{Chiorescu04}
\bibinfo{author}{Chiorescu, I.} \emph{et~al.}
\newblock \bibinfo{title}{Coherent dynamics of a flux qubit coupled to a
  harmonic oscillator}.
\newblock \emph{\bibinfo{journal}{Nature}} \textbf{\bibinfo{volume}{431}},
  \bibinfo{pages}{159--162} (\bibinfo{year}{2004}).

\bibitem{Ilichev03}
\bibinfo{author}{Il'ichev, E.} \emph{et~al.}
\newblock \bibinfo{title}{Continuous monitoring of rabi oscillations in a
  josephson flux qubit}.
\newblock \emph{\bibinfo{journal}{Phys. Rev. Lett.}}
  \textbf{\bibinfo{volume}{91}}, \bibinfo{pages}{097906}
  (\bibinfo{year}{2003}).

\bibitem{Blais04}
\bibinfo{author}{Blais, A.}, \bibinfo{author}{Huang, R.},
  \bibinfo{author}{Wallraff, A.}, \bibinfo{author}{Girvin, S.~M.} \&
  \bibinfo{author}{Schoelkopf, R.~J.}
\newblock \bibinfo{title}{Cavity quantum electrodynamics for superconducting
  electrical circuits: An architecture for quantum computation}.
\newblock \emph{\bibinfo{journal}{Phys. Rev. A}} \textbf{\bibinfo{volume}{69}},
  \bibinfo{pages}{062320} (\bibinfo{year}{2004}).

\bibitem{Martin04}
\bibinfo{author}{Martin, I.}, \bibinfo{author}{Shnirman, A.},
  \bibinfo{author}{Tian, L.} \& \bibinfo{author}{Zoller, P.}
\newblock \bibinfo{title}{Ground state cooling of mechanical resonators}.
\newblock \emph{\bibinfo{journal}{Phys. Rev. B}} \textbf{\bibinfo{volume}{69}},
  \bibinfo{pages}{125339} (\bibinfo{year}{2004}).

\bibitem{Rabl04}
\bibinfo{author}{Rabl, P.}, \bibinfo{author}{Shnirman, A.} \&
  \bibinfo{author}{Zoller, P.}
\newblock \bibinfo{title}{Generation of squeezed states of nanomechanical
  resonators by reservoir engineering}.
\newblock \emph{\bibinfo{journal}{Phys. Rev. B}} \textbf{\bibinfo{volume}{70}},
  \bibinfo{pages}{205304} (\bibinfo{year}{2004}).

\bibitem{Niskanen07}
\bibinfo{author}{Niskanen, A.~O.}, \bibinfo{author}{Nakamura, Y.} \&
  \bibinfo{author}{Pekola, J.~P.}
\newblock \bibinfo{title}{Information entropic superconducting microcooler}
  (\bibinfo{year}{2007}).
\newblock \bibinfo{note}{ArXiv:0704.0845v1}.

\bibitem{Hauss07}
\bibinfo{author}{Hauss, J.}, \bibinfo{author}{Fedorov, A.},
  \bibinfo{author}{Hutter, C.}, \bibinfo{author}{Shnirman, A.} \&
  \bibinfo{author}{Sch{\"o}n, G.}
\newblock \bibinfo{title}{Single-qubit lasing and cooling at the rabi
  frequency}  (\bibinfo{year}{2007}).
\newblock \bibinfo{note}{ArXiv:cond-mat/0701041v5}.

\bibitem{Wineland92}
\bibinfo{author}{Wineland, D.~J.}, \bibinfo{author}{Dalibard, J.} \&
  \bibinfo{author}{Cohen-Tannouji, C.}
\newblock \bibinfo{title}{Sisyphus cooling a bound atom}.
\newblock \emph{\bibinfo{journal}{J. Opt. Soc.}} \textbf{\bibinfo{volume}{B9}},
  \bibinfo{pages}{32--42} (\bibinfo{year}{1992}).

\bibitem{Mooij99}
\bibinfo{author}{Mooij, J.~E.} \emph{et~al.}
\newblock \bibinfo{title}{Josephson persistent-current qubit}.
\newblock \emph{\bibinfo{journal}{Science}} \textbf{\bibinfo{volume}{285}},
  \bibinfo{pages}{1036--1039} (\bibinfo{year}{1999}).

\bibitem{Quan07}
\bibinfo{author}{Quan, H.~T.}, \bibinfo{author}{Liu, Y.~X.},
  \bibinfo{author}{Sun, C.~P.} \& \bibinfo{author}{Nori, F.}
\newblock \bibinfo{title}{Method for direct observation of coherent quantum
  oscillations in a superconducting phase qubit}.
\newblock \emph{\bibinfo{journal}{Phys. Rev. E}} \textbf{\bibinfo{volume}{76}},
  \bibinfo{pages}{031105} (\bibinfo{year}{2007}).

\bibitem{Greenberg02a}
\bibinfo{author}{Greenberg, Y.~S.} \emph{et~al.}
\newblock \bibinfo{title}{Method for direct observation of coherent quantum
  oscillations in a superconducting phase qubit}.
\newblock \emph{\bibinfo{journal}{Phys. Rev. B}} \textbf{\bibinfo{volume}{66}},
  \bibinfo{pages}{224511} (\bibinfo{year}{2002}).

\bibitem{Ithier05}
\bibinfo{author}{Ithier, G.} \emph{et~al.}
\newblock \bibinfo{title}{Decoherence in a superconducting quantum bit
  circuit}.
\newblock \emph{\bibinfo{journal}{Phys. Rev. B}} \textbf{\bibinfo{volume}{72}},
  \bibinfo{pages}{134519} (\bibinfo{year}{2005}).

\bibitem{Gardiner}
\bibinfo{author}{Gardiner, C.~W.} \& \bibinfo{author}{Zoller, P.}
\newblock \emph{\bibinfo{title}{Quantum noise}} (\bibinfo{publisher}{Springer},
  \bibinfo{year}{2004}), \bibinfo{edition}{3-d} edn.

\bibitem{naik-2006-443}
\bibinfo{author}{Naik, A.} \emph{et~al.}
\newblock \bibinfo{title}{Cooling a nanomechanical resonator with quantum
  back-action}.
\newblock \emph{\bibinfo{journal}{Nature}} \textbf{\bibinfo{volume}{443}},
  \bibinfo{pages}{193--196} (\bibinfo{year}{2006}).

\bibitem{blanter-2004-93}
\bibinfo{author}{Blanter, Y.~M.}, \bibinfo{author}{Usmani, O.} \&
  \bibinfo{author}{Nazarov, Y.~V.}
\newblock \bibinfo{title}{Single-electron tunneling with strong mechanical
  feedback}.
\newblock \emph{\bibinfo{journal}{Phys. Rev. Lett.}}
  \textbf{\bibinfo{volume}{93}}, \bibinfo{pages}{136802}
  (\bibinfo{year}{2004}).

\bibitem{blencowe-2005-7}
\bibinfo{author}{Blencowe, M.~P.}, \bibinfo{author}{Imbers, J.} \&
  \bibinfo{author}{Armour, A.~D.}
\newblock \bibinfo{title}{Dynamics of a nanomechanical resonator coupled to a
  superconducting single-electron transistor}.
\newblock \emph{\bibinfo{journal}{New J. Phys.}} \textbf{\bibinfo{volume}{7}},
  \bibinfo{pages}{236} (\bibinfo{year}{2005}).

\bibitem{bennett-2006-74}
\bibinfo{author}{Bennett, S.~D.} \& \bibinfo{author}{Clerk, A.~A.}
\newblock \bibinfo{title}{Laser-like instabilities in quantum
  nano-electromechanical systems}.
\newblock \emph{\bibinfo{journal}{Phys. Rev. B}} \textbf{\bibinfo{volume}{74}},
  \bibinfo{pages}{201301} (\bibinfo{year}{2006}).

\bibitem{MRFM}
\bibinfo{author}{Sidles, J.~A.} \emph{et~al.}
\newblock \bibinfo{title}{Magnetic resonance force microscopy}.
\newblock \emph{\bibinfo{journal}{Rev. Mod. Phys.}}
  \textbf{\bibinfo{volume}{67}}, \bibinfo{pages}{249--265}
  (\bibinfo{year}{1995}).

\end{thebibliography}
\end{document}